\documentclass{aa}
\usepackage{epsfig,amsmath}

\def\equaldef{\buildrel {def} \over {=}}
\begin{document}

\title{Interacting galaxies and cosmological parameters}
\author{H. Reboul\inst{} \and J.--P. Cordoni\inst{}}  
\offprints{H. Reboul, \\ \email{reboul@graal.univ-montp2.fr}}

\institute{{UMR 5024, CNRS--Universit\'e
Montpellier 2, GRAAL, CC 72, 
         34095 Montpellier Cedex 5,
	 France}
  }

\titlerunning{Interacting galaxies and cosmology} 
        
\date{Version 6 -  / Received  / Accepted }

\abstract{We propose a (physical)-geometrical method to
measure $\Omega_{m\circ}$ and $\Omega_{\Lambda\circ}$, the present rates
of the density cosmological parameters for a 
Friedmann-Lema\^{\i}tre
universe. The distribution of linear separations between two interacting 
galaxies,
when both of them undergo a first massive starburst,  is used as a standard of length. 
Statistical properties of the linear separations of such pairs of ``interactivated'' 
galaxies 
are estimated from the data in the
Two Degree Field Galaxy Redshift Survey.
Synthetic samples of interactivated pairs are generated with random orientations 
 and a likely distribution of redshifts.  The resolution of 
 the inverse problem provides the probability densities of the retrieved cosmological parameters.  
 The accuracies that can be achieved by that method on $\Omega_{m\circ}$ and $\Omega_{\Lambda\circ}$
  are computed
 depending on the size of ongoing real samples. Observational prospects are 
 investigated as the foreseeable surface densities on the sky and magnitudes of 
 those objects.

}

\maketitle

\section{Introduction}

Variation in the scale factor $R(t)$ of a Friedmann-Lema\^{\i}tre (FL) universe 
with cosmic time $t$ 
 affects 
the observable relations $m \longleftrightarrow z_c$ and $\theta \longleftrightarrow z_c$
between apparent magnitude $m$ and angular size
$\theta$ versus the cosmological redshift $z_c$ of standard sources. When possible,
a solution to the 
inverse problem 
may then supply
the whole story of $R(t)$ and the spatial curvature. \\

The $m \longleftrightarrow z_c$ relation provided  the first 
estimation of the expansion  rate
$ H \equaldef \dot R / R$  (Lema\^{\i}tre \cite {Lem27}), a long time ago. 
Much more recently, supernovae SNIa (Riess et al. \cite {Riess98}, Perlmutter et al. 
\cite {Perlm99})
were   
the standard candles that accredited -- with the help of the angular power spectrum of the 
anisotropy for the 
Cosmic Microwave Background Radiation (CMBR) -- the so-called ``concordance model'' 
in which the  density parameters 
for cold matter ($\Omega_m \equaldef 8 \pi G\rho_m / 3 H^2$) and for cosmological constant
($\Omega_\Lambda \equaldef \Lambda c^2 / 3 H^2$) have the present (index${}_\circ$)
values
$\Omega_{m\circ} \sim 1/3$ and $\Omega_{\Lambda\circ} \sim 2/3$.
 All this revived
 a dominant $\Omega_{\Lambda\circ}$ universe, after Lema\^{\i}tre (\cite {Lem27}, 
\cite{Lem31}).  But as pointed out by Blanchard 
et al (\cite{Blan03}), that concordance is not entirely free from
weak hypotheses,
and those authors argued that the previously dominant Einstein-de-Sitter model 
($\Omega_{m} = 1$ and $\Omega_{\Lambda} = 0$) was still not
excluded by available data.\\

The case for $\Omega_{\Lambda\circ}$ is important. That parameter 
is not only determinant for the geometrical age of the universe and 
for the evolution of large structures but, 
in the FL equations
 on the scale factor $R(t)$, 
the geometrical cosmological constant $\Lambda$ 
may be, at least formally and partly or totally, exchanged
 with a physical
``vacuum energy'', a perfect and Lorentz invariant fluid of equation of state $p=w \rho c^2$
with $w_\Lambda = -1$ and 
$\rho_\Lambda = \Lambda c^2 / 4 \pi G$ (Lema\^{\i}tre \cite {Lem34}) that this author  judged 
to be {\it ``essentially the meaning of the cosmical constant''}. 
And the cosmological tests that detected 
$\Omega_{\Lambda\circ} \not = 0$ may also be used to constrain the $w\not = -1$ or 
$w\not = constant$ of more elusive fluids like dark energy or 
quintessence.\\

To that purpose the $\theta \longleftrightarrow z_c$ relation 
is also a promising cosmological test that was first investigated by Tolman (\cite {Tol30}). 
If an object has  a projected linear separation $PS$ on 
the plane of the sky at the time of emission $t_e$, 
the radial motion of received photons leads to an observed angular size:
\begin{equation}
\theta_\circ (z_c) = \frac{PS \ (1+z_c)}{d} \equaldef \frac{PS}{d_A} \ \ .
\end{equation}
In the above expression, $d$ is the ``metric'' or ``comoving tranverse'' or 
``proper motion'' 
``distance'' of the source
and $d_A$ is its ``angular size--distance''.  The expressions for $d$ are 
obtained by intregration of the radial light's movement from the source to the observer
($d_-$, $d_f$,  $d_+$ respectively are for negative, null, and positive curvatures
 of space, 
$\Omega_r$ is the density parameter of radiation, and
 $\Omega_k \equaldef -kc^2/R^2H^2$ is the reduced 
curvature of space related to other $\Omega_i$ by
 $\Omega_m + \Omega_r + \Omega_\Lambda + \Omega_k=1$) 
:
\begin{equation}d_- = \frac{c}{H_\circ \mid \Omega_{k \circ} \mid^{\frac{1}{2}}} \
\sinh \left [ \mid \Omega_{k \circ} \mid^{\frac{1}{2}} F(z_c) \right ] 
\ \ \ (H^3 \ \rm{space})\end{equation}
\begin{equation}d_f = \frac{c}{H_\circ} \ F(z_c) \ \ \ (E^3 \ \rm{space})\end{equation}
\begin{equation}d_+ = \frac{c}{H_\circ  \mid \Omega_{k \circ} \mid^{\frac{1}{2}}} \
\sin \left [ \mid \Omega_{k \circ} \mid^{\frac{1}{2}} F(z_c) \right ] 
\ \ \ (S^3 \ \rm{space})\end{equation}
with
\begin{equation}
\begin{array}{l}
\displaystyle{
F(z_c)=\int_0^{z_c} [\Omega_{r\circ} (1+x)^4 + \Omega_{m\circ} (1+x)^3} \ \ \ \ \ \ \ \ \ \ \ \ \\
 \ \ \ \ \ \ \ \ \ \ \ \ \ 
  \ \ \ \ \ \ \ \ \ \ \ \ \ + 
  \Omega_{k\circ} (1+x)^2 + \Omega_{\Lambda \circ} ]^{- \frac{1}{2}} \ dx \ \ \ .
\end{array}
\end{equation}
We note that the $\theta_\circ \longleftrightarrow z_c$ relation 
is not directly linked to the projected separation $PS$ but to the product 
$H_\circ \cdot PS$, if
the observational determination of real $PS$  uses -- and is then inversely proportional to --
the rate of $H_\circ$, i.e. when distances are deduced from redshifts 
and not directly from indicators.\\

\begin{figure}
\hspace{-0.9cm}
\psfig{figure=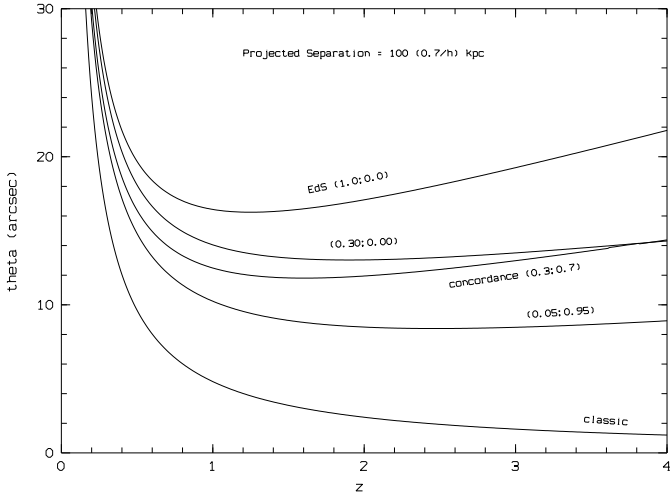,width=7.0cm,angle=-90}
\caption[]{Sensitivity of the $\theta_\circ \longleftrightarrow z_c$ 
relation to $\Omega_{m\circ}$ and $\Omega_{\Lambda\circ}$. }
\label{thetadez}
\end{figure}

The discriminating power of the 
$\theta_\circ \longleftrightarrow z_c$
relation versus some sets of cosmological parameters is displayed in
Fig.~\ref{thetadez}. For currently favoured cosmological models, $\theta_\circ$ remains 
greater than a minimum value :
\begin{equation}\theta_\circ(\arcsec) > \frac{PS(kpc)}{10} \ \ \ . \end{equation}

Attempts to constrain the cosmological parameters with the 
$\theta_\circ \longleftrightarrow z_c$ relation have been performed. 
First conclusive results with radio-sources have been suggested by
Kellermann (\cite {Kell93}) for the deceleration parameter $q_\circ$
($q= \Omega_r + \Omega_m / 2 -\Omega_\Lambda$ and $q_\circ \approx \Omega_{m\circ} / 2
 -\Omega_{\Lambda\circ}$ 
in our expanded universe).
As reviewed by Gurvits et al. (\cite {Gurv99}), the large radio structures had supplied 
inconclusive or paradoxical data : classical $\theta_\circ \propto 1/z$ or 
even $\theta_\circ \sim$ constant! Those authors
did focus on milliarcsec 
compact radio sources, which are presumably physically very young so not very related 
to or affected by the cosmic evolution  of the intergalactic medium. They 
derived a constraint in that way  on the deceleration parameter $q_\circ = 0.21 \pm 0.30$. 
Guerra et al. (\cite{Guerra00}) have obtained wide contours in the 
($\Omega_{m\circ}$, $\Omega_{\Lambda\circ}$) plane with 20 powerful 
double-lobed radio galaxies as 
yardsticks.
Lima \& Alcaniz (\cite{Lima02}) and  Chen \& Ratra (\cite{Chen03}), using 
the data of Gurvits et al. (\cite {Gurv99}), have both derived wide constraints 
on the densities parameters
and also on the index in the expression of the potential 
of the dark energy scalar 
field, which could challenge $\Lambda$.
Zhu \& Fujimoto (\cite{Zhu02}) with the data of Gurvits et al. (\cite {Gurv99}),
Zhu et al. (\cite{Zhu04b}) using the data of Guerra et al. (\cite{Guerra00}),
and
 Zhu \& Fujimoto (\cite{Zhu04}) also investigated the 
$\theta_\circ \longleftrightarrow z_c$ relation to constrain the $w$ parameter 
of dark energy
and the free parameters 
of non-standard cosmologies.\\

The main problems encountered with astrophysical objects -- or non-interacting pairs -- in the
cosmological utilization of their $\theta_\circ(z_c)$ relation are:
\begin{itemize}
\item i)  the statistical evolution of
linear size with cosmic time (and then with $z_c$) as already mentioned for radio sources. 
This is well known for
the clusters of galaxies whose relaxation time is comparable to the Hubble time 
($H_\circ ^{-1}$) and which 
are still accreting material in the central parts of superclusters.
The more recent discovery that the bulk of galaxies are the result of multiple merging processes
seems to exclude them for that purpose. As the intergalactic medium IGM is also evolving 
with cosmic
time, the selection of very young (unmerged) galaxies does not seem to be a solution.
\item ii) measurement biases due to fuzzy intrinsic photometric profiles of objects 
(galaxies, clusters of galaxies) and to the fast decrease in surface brightness 
with redshift.
\item iii) redshifts 2 to 3 have to be reached to disentangle the partial degeneracy between 
$\Omega_{m\circ}$ 
and $\Omega_{\Lambda\circ}$.
\end{itemize}

With the purpose of avoiding the most important part of those drawbacks, 
our idea is to replace 
the standard objects
by pairs of bright related objects. Physically this consists in finding pairs of objects 
displaying a special feature because they are at a characteristic physical
distance from each other. Replacing diffuse objects by pairs of point-like 
(or relatively well ``picked") sources 
removes observational 
bias in the measure of $\theta_\circ$. If the physical process that causes the special feature 
is not sensitive to the cosmic evolution, the main drawback of the method is removed. 
If the objects furthermore have 
strong emission lines, measuring their redshift becomes obvious and the 
$\theta_\circ \longleftrightarrow z_c$ relation
may then become an efficient way to measure cosmological parameters.\\

We long ago proposed to use this method with {\it ``interactivating AGNs"}
or {\it ``really double QSOs''}  (Reboul et al. \cite {reb85}). 
At that time those objects had just been discovered (Djorgovski et al. \cite {Djorg87}), and
we considered a 
very wide field 
survey of interactivating double QSO at a limiting magnitude of $\sim 20$. 
We began a systematic search for these objects through a primary selection
by colour criteria on Schmidt plates (Reboul et al. \cite {reb87}; 
Vanderriest \& Reboul \cite {Van91}; Reboul et al. \cite
{reb96}). (Another motivation of that search was to look for gravitational mirages).\\

True interactivated
pairs of AGNs -- essentially QSOs or Seyferts -- are very
uncommon: 14 cases of binary QSOs in the 11th V\'eron and V\'eron catalogue (\cite{Ver03}).
But in fact real pairs of QSOs are the extreme avatar of the more common {\it ``interactivation 
of galaxies''} by which we mean the mutual transformation of
two encountering galaxies into a temporary pair of active objects (starbursts or sometimes AGNs).\\

The tidal 
deformations of encountering galaxies, their occasional merging and the resulting 
stellar streams in the merged object are now depicted fully by numerical simulations
ever since the pioneering works of Toomre and Toomre (\cite {Toomre72}). But the whole 
dissipative process by which a  close encounter of galaxies triggers 
observable massive
starbursts and (sometimes) true AGNs is extremely complex and
extends over a huge dynamical range of distances and densities.\\

The complete modeling, including induced starbursts, is more recent. 
Barnes \& Hernquist (\cite {Barnes91}) have proved the rapid fall of 
gas towards nuclei in a merger. Mihos \& Hernquist (\cite {Mihos94}) computed the evolution
of the global star-formation rate (SFR) in galaxy merger events. Their Fig. 2, 
like the Fig. 1 of
Springel \& Hernquist (\cite {Spri05}),
clearly demonstrates the two episodes of starburst 
in a merging encounter.\\

The primary starburst is induced by the first approach of the two galaxies. 
In the standard scenario, the dynamical friction transforms a quasi parabolic 
(minimal relative velocity 
and then maximum tidal efficiency) initial orbit before periapse
into a one-tour quasi-elliptic one. The second and closer approach 
is much more dissipative and soon evolves in the merging.\\

In fact it is the second step that has been mainly studied in recent years. 
This intense, condensed, short, and dusty starburst is the likely source of extreme objects 
like ultra-luminous
infrared galaxies (see Sanders \& Mirabel, \cite {San96} for a review). \\

On the contrary, we expect the primary 
starburst to be the generator
of yardsticks, through the combination of its luminosity 
curve and the first part of the bouncing relative 
orbit. The first bounce also has the qualities of large separations 
 and well-defined central profiles for the two galaxies supplying easy measure of angular
 separation $\theta_\circ$. \\

The main purpose of this paper is to quantify the expected performances
of such a 
method to constrain
cosmological parameters through observations of primary interactivating galaxies.

\section{Low-redshift sample}
There is no available homogeneous sample of well-defined pairs of interactivated galaxies.
Our own samples of FRV (Fringant et al. \cite {frv83}, Vanderriest \& Reboul \cite{Van91},
Reboul \& Vanderriest \cite {reb02} and references 
herein) 
sources were those that revealed to us
a characteristic
distance for interactivated galaxies and the narrow photometric 
profile of central starbursts (FWHM typically less than 500 pc). 
But, initially intended to find true ``mirages'' 
(gravitational lenses), those samples were  limited from the start to 
angular separations less than  $\sim 10\arcsec$ and are then presumably biased 
in favour of mid-evolved 
(close to merging) secondary starburst systems and in disfavour of long bouncing 
primary interactivation pairs.
So we looked for another 
source to help  estimate  the statistical 
properties of the 
geometrical parameters for interactivated galaxies. \\

\begin{figure}
\vspace{0.3cm}
\hspace{-0.5cm}
\psfig{figure=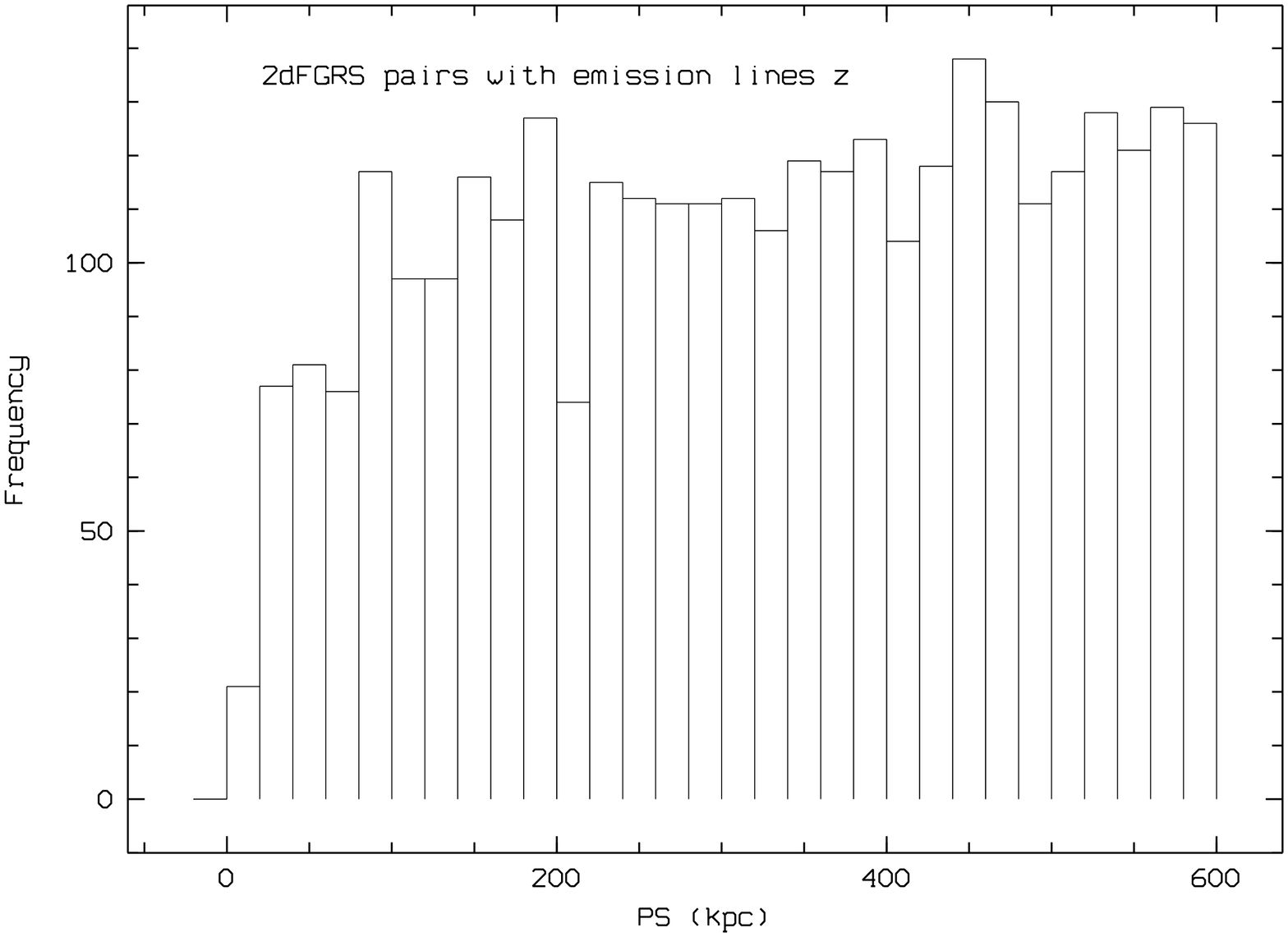,width=6.4cm,angle=-90}
\caption[]{Histogram of the distribution of projected separations
for the 3239 pairs in the 2dFGRS selected by : 
i) redshifts from emission lines and greater than 0.001, ii) angular separation less than 10\arcmin.
The concordance model is assumed.}
\vspace{0.cm}
\label{histo3239}
\end{figure}

The release of the 2dFGRS Final Data Spectroscopic 
Catalogue 
(Colles et al. \cite{Colless03})
was an opportunity. We performed a systematic search of pairs among its  245 591 entries.
We display (Fig.~\ref{histo3239}) the histogram for the distribution of projected separations
for all the pairs of objects in  the 2dFGRS catalogue that have redshifts measured 
by emission lines (and greater than 0.001) and angular separations less than 10\arcmin.
A concordance $\Lambda$CDM model was assumed: 
$h_\circ=0.7, \ \Omega_{m\circ} = 0.3, \ \Omega_{\Lambda\circ} = 0.7$ 
($H_\circ \equaldef 100 \ h_\circ \ $ km s$^{-1}$ Mpc$^{-1}$).
At those short distances $\Omega_{\Lambda \circ}$ is 
quite inefficient. We checked that the cut-off in angular separation does not 
significantly affect the histogram of projected separation in the displayed range.\\ 

We retained the 
following criteria for  selection of 
the interactived candidates:
\begin{itemize}
\item angular separations $\theta_\circ \leq 10 \arcmin$
\item magnitude difference: $\left | B_1 - B_2 \right | < 2$
\item redshift of the two members of the pair measured by their emission lines 
(which is an a priori sign of a high ratio of
emission lines versus continuum)
\item number of identified emission lines $N_{el} \geq 5$ for the two members
\item heliocentric redshift $z\ge 0.001$ to get observed $z \approx z_c$ in avoiding
 too high an interference of 
Doppler-Fizeau redshifts due 
to local motions of the centre of mass of interactivating galaxies
\item relative radial velocities with cosmological correction
\begin{equation}\Delta v_r = \frac{2c\mid z_2-z_1\mid}{2+z_1 + z_2} < 75 \ \rm{km} \ \rm{s}^{-1}\end{equation}
\item projected separations $PS < 300 \ (0.7/h_\circ)$ kpc  computed with the 
``concordance'' FL model.
\end{itemize}

\begin{figure}
\vspace{0.cm}
\hspace{-0.9cm}
\psfig{figure=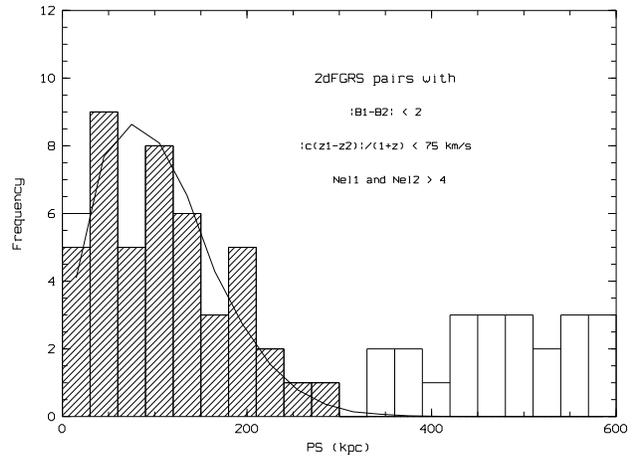,width=6.4cm,angle=-90}
\vspace{0.cm}
\caption[]{Selection of the sub-sample of interactivated galaxies 
candidates in the 2dFGRS. The zone of the 45 retained pairs with $PS < 300 \ (0.7/h_\circ)$ kpc 
is hashed. The white case
in the first bin is for the rejected pair in the arms of NGC 4517. 
The solid line displays the Poissonian distribution used for simulations.}
\vspace{0.3cm}
\label{histo68}
\end{figure}

The final selection is displayed in Fig.~\ref{histo68}.
Purged of two redundances (caused by a triplet) the ``final'' sample contains 68 pairs. 
As shown in Fig.~\ref{histo68}, a population of 
46 pairs with $PS < 300 \ (0.7/h_\circ)$ kpc seems separable from the general background of more random 
associations.\\

A close inspection of those 46 pairs on DSS images revealed that one of them (300591-- 300593) 
is probably formed
by two HII regions in 
the complex of the perturbed (merged ?) galaxy NGC 4517. We removed that pair.\\


The Multi-Object-Spectroscopy (MOS) with addressable fibres may induce a selection bias 
against close pairs through the mechanical width of ``buttons''
that attach fibres on the field plate,
practically 33\arcsec \hskip 1 mm for the 2dF spectrograph. This drawback may be 
compensated for by a pertinent redundance of exposures. That bias may be evaluated 
(Mathew Colless, priv. com) when comparing
photometric and spectroscopic catalogues of the 2dF. This inspection showed that the distribution 
of the number of pairs in angular separations are  related well in the  two catalogues above, 
as the number of 
pairs in the photometric catalogue are 30 \% higher than in the spectroscopic one. 
That works, except for
pairs closer than 18\arcsec \hskip 1 mm for which the ratio is higher. As there are
 only three such close pairs in our selected sample, we may estimate that this does not induce 
 a noticeable bias  in our estimated  distribution of projected separations.\\

We then chose to extract the parameters of the distribution of $PS$ for 
interactivated galaxies
from the sub-sample of 45 pairs with $PS < 300 (0.7/h_\circ)$ kpc.
Absolute magnitudes of the 90 objects range from -15.1 to -20.7 with -19.2 
for the magnitude of mean luminosity and redshifts from 0.009 to 0.108 with a mean of
0.052.
We fitted the histogram of $PS$ in Fig.~\ref{histo68} with a 
Poissonian probability law (a first attempt 
with a 
lognormal law was less satisfactory). The mean -- and variance -- of the Poissonian 
fitting is 
$105.3 \ (0.7/h_\circ)$ kpc .\\

We do not 
claim  here  to achieve a real measurement of the distribution of 
the projected separations of
local interactivating galaxies: in the best 
case we got an estimation (presumably a majorant) of the relative dispersion of 
the $PS$. And there lies all  we need to qualify the method.
We note that the precise parameters of the real population will
be updated on a larger sample with incoming data by the statistical study of the low redshift 
 pairs that are spectroscopically confirmed.\\

At least, we tried to evaluate the orbital parameters of such encounters. A simple modeling
with a Keplerian orbit between the first and second approaches and with an evolution of the SFR
that is similar to that of Springel \& Hernquist (\cite{Spri05})  seems to favour  massive galaxies 
($M_1 + M_2 \sim 10^{12}$ M$_\odot$) for the 
selected population.
Fitting both the observed distribution of separations and that of relative radial 
velocities $\Delta v_r$ would  indicate longer ($\times 2$ 
to $\times 3$) starbursts.\\

As explained above, we use that distribution of local projected separations
to generate the synthetic 
samples, which includes the hypothesis that the statistical properties of the geometrical
parameters of interactivations are independent of cosmic time. Reliance 
on that assumption 
is theoretically based on the consideration that the {\it primum movens}  of 
both the interactivation -- starburst -- process and real separations is the -- a priori constant --
gravitational interaction.  It is also founded
on the fact that the $\sim 2.5 \ 10^5$ 2dFGRS galaxies, from which 
our sample has been selected, have a  wide dynamic 
of individual characteristics like masses and  gas fractions. 
Then a statistical evolution of the characteristics of individual galaxies with redshift
could have no first order effect on the linear separations of scarce interactivated pairs.
The selection of primary interactivations is also an asset: those pairs are preferably 
constituted with gas-rich galaxies which are still quite free of strong merger experience.
At any rate, numerical 
simulations would be the best way of quantifying how sensitive  the distribution of separations 
is to parameters like mass, gas fraction, and gas properties of galaxies 
and then to estimate -- and possibly correct -- 
a redshift dependence.

\section{Synthetic samples}

 \subsection{Distribution in orientation}

If $i$ is the inclination of the pair on the line of sight, the real linear separation $LS$ is 
related to
 the projected separation $PS$ by $LS = PS / \sin i$. 
If $i$ is not an easily observed parameter,
the natural hypothesis for an isotropic distribution of pair orientation  makes 
the set of 
 possible directions homeomorphic 
 to a Euclidean 2-sphere,
 and a simple integration on that sphere supplies the mean values:
$<i> = 1$ rad  and $<\sin i> = \pi/4$.
 It is worth noting the latter value ($\sim 0.8$) of this projection factor, 
  since it will explain why the unavailability of 
 $i$ in the observations 
 will not add a strong dispersion.

\subsection{Distribution in linear separation}

The expectation of the product of two independent random variables is the product of their 
expectations. Then
the distribution of linear separations for the pairs of interactivated galaxies would 
have an expectation $105.3 \ / \ (\pi/4) = 134.1 \ (0,7/h_\circ)$ kpc.\\

There are two reasons for the dispersion of $PS$: linear separations and
random orientations.
The latter dispersion is that of  $\sin i$.  It has a 
standard 
deviation $\sigma
\sim 0.22$ or
a ``relative dispersion'' (standard deviation to mean ratio) 
of $\sim 0.22/(\pi/4) \approx 0.28$. 
That of the $PS$ of the 45 pairs extracted from the  2dFGRS is much greater : $72 / 109 
\approx 0.66$.
Then the dispersion of the real sample is mainly due to the physical dispersion of 
linear separations, and the random inclination 
does not
greatly affect the potentiality of the method. \\

We generated the linear separations in the mock samples of interactivated galaxies 
as a Poissonian distribution with expectation $134.1 \ (0,7/h_\circ)$ kpc before applying 
the projection effect of random orientation (previous subsection).

\subsection{Distribution in redshift}
The parentage between nuclear starbursts and true active nuclei, the similarity 
in their observing techniques, and the lack of deep samples of interactivated
galaxies, all made it 
seems natural to use the distribution of redshift for a homogeneous
 sample of quasars. \\
\begin{figure}
\vspace{0.3cm}
\hspace{-0.5cm}
\psfig{figure=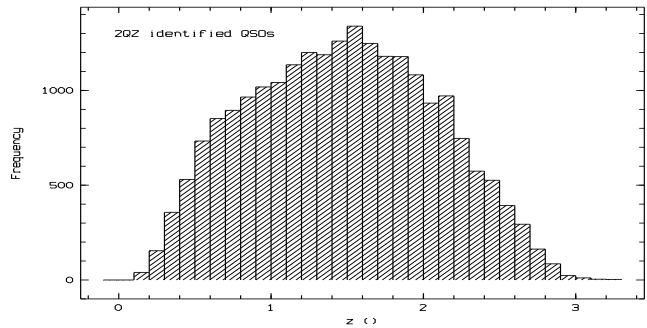,width=4.6cm,height=9.4cm,angle=-90}
\caption[]{Redshifts of QSOs in the 2QZ survey. We selected the 22 122 objects
with only the  label ``QSO ''.
This distribution of redshifts has been applied to 
the synthetic samples of interactivating galaxies}
\vspace{0.3cm}
\label{histoz}
\end{figure}
 
 We chose the two-degree Field QSO redshift survey 
 (2QZ) (Croom et al. \cite{Croom04}) in which we 
 selected those 22 122 objects
 with the label ``QSO ''. The histogram of redshifts is displayed in Fig.~\ref{histoz}.
 In our synthetic process each pair then received 
 a random redshift from that data base.
All the random numbers and distributions above were generated with subroutines imported from 
``Numerical Recipes in Fortran''
(imported from Press et al. \cite{Pre92}).

\subsection{Distribution in $\Omega_{i\circ}$}
With the purpose of estimating the  inhomogeneity in the sensitivity of the method through 
the credible part of the $\Omega_{i\circ}$ field, 
we applied the whole procedure
to a small set of 
tentative couples $(\Omega_{m\circ}, \ \Omega_{\Lambda\circ}$).
Then mock samples of ($\theta_\circ$, $z$) were generated through the Monte-Carlo 
method  described above and with
the general cosmological 
relations reviewed in Sect. 1.

\section{Retrieving $\Omega_{i\circ}$}
Retrieving $\Omega_{m\circ}$ and $\Omega_{\Lambda\circ}$ from the synthetic samples
was  solved by the Levenberg-Marquardt (LM) technique 
(routines in ``Numerical Recipes'') which seemed well-suited to the  non-linear 
and entangled inverse problem.\\

\begin{figure}
\vspace{0.0cm}
\hspace{0.0cm}
\psfig{figure=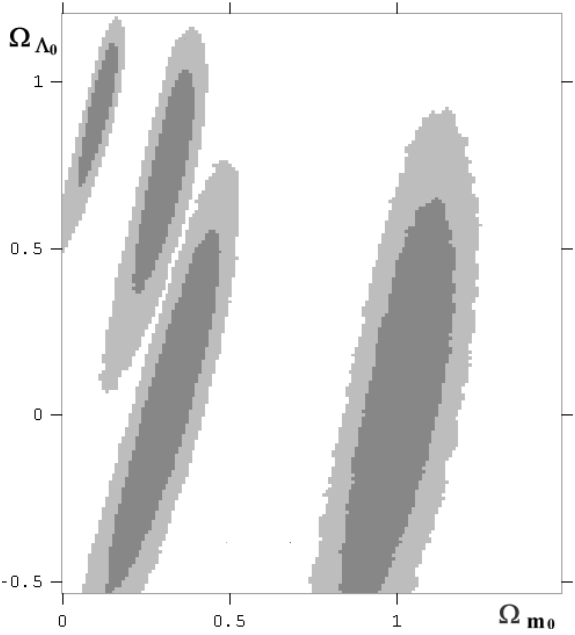,width=7.6cm}
\vspace{0.0cm}
\caption[]{Normalized probability densities of retrieved cosmological parameters with 
1000 pairs of interactivated galaxies. The 68\% and 95\% confidence levels appear 
in shaded and darker shaded areas for the four sets 
of ($\Omega_{m\circ}$, $\Omega_{\Lambda\circ}$): 
(1.0, 0.0), (0.3, 0.0), (0.3, 0.7), and (0.1, 0.9).}
\vspace{0.cm}
\label{omegas}
\end{figure}
\begin{figure}
\vspace{0.1cm}
\hspace{-0.5cm}
\psfig{figure=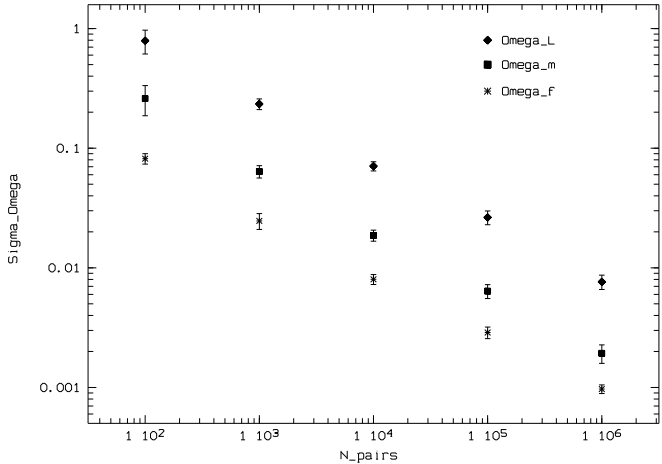,width=6.6cm,angle=-90}
\vspace{0.5cm}
\caption[]{Internal precision ($\sigma$) on $\Omega_{m\circ}$ and $\Omega_{\Lambda\circ}$ 
versus number of pairs and for the ``concordance'' model,
also shown
 with $\Omega_{f\circ}$ the precision with the flat space external condition 
 $\Omega_{m\circ} + \Omega_{\Lambda\circ} =1$.}
\vspace{0.cm}
\label{sigmas}
\end{figure}
Figure ~\ref{omegas} displays the potentiality of the method through the plausible zone of
the ($\Omega_{m\circ}, \ \Omega_{\Lambda\circ}$) field and only for 1000 pairs. We chose 
these four combinations:\\ 
(1.0, 0.0), (0.3, 0.0), (0.3, 0.7) and (0.1, 0.9).
We assumed that all redshifts were known precisely. We checked the 
inversibility by   applying that method to samples generated 
with zero dispersion in linear projected separations.\\

We summarise in Fig.~\ref{sigmas} the standard deviations on $\Omega_{m\circ}$ and 
$\Omega_{\Lambda\circ}$ 
resulting from simulations ranging from $10^2$ to $10^6$ pairs. If an external condition 
is added to the sum $\Omega_{m\circ} + \Omega_{\Lambda\circ}$ 
(as with CMBR), the accuracy of 
the method is obviously enhanced.
The standard deviations $\sigma_i$ of the fitted parameters  all display a 
nominal decrease:
 $\sigma_i \propto 1/ \sqrt {N\_ pairs}$.
As a matter of fact those 
accuracies are only internal 
to the method.

\section{Observational prospects}
\subsection{Foreseeable data}
As a surface of constant cosmic time of emisson $t_e$ is isometric 
to a Euclidean 2-sphere of radius $R(t_e)r$,
the elementary volume in a 1-steradian pencil  and for sources 
emitting in the cosmic time interval
$dt_e$ (thickness $dl$) is 
\begin{equation}dV = d_A^2 \ dl = d_A^2 \ c \ dt_e\ .\end{equation}
If $n(z_c)$ is the density at redshift $z_c$, the number of sources 
per steradian in the range $\left[z_c , z_c+ dz_c\right]$ 
may be expressed as:
\begin{equation}dN =n(z_c) \ d_A^2\frac{c}{H(z_c) (1+z_c)} \ dz_c\ .\end{equation}
FL equations lead to 
\begin{equation}\frac{H}{H_\circ} = \left[\Omega_{m\circ} (1+z_c)^3 + 1 - 
\Omega_{m\circ} \right]^{{1\over 2}}\end{equation}
for a flat $\Lambda$CDM universe.
Assuming $n(z_c) = f_\circ (1+z_c)^m$ ($m=3$ for a constant comoving density of sources),
the total number of sources per steradian 
in the interval 
$\left[z_{c1},z_{c2} \right]$ may be deduced:
\begin{equation}
\begin{array}{l}
\displaystyle{
{N} = f_\circ \frac{c^3}{H_\circ^3}  \int_{z_{c1}}^{z_{c2}}}
[\Omega_{m\circ} (1+x)^3 + 1 - \Omega_{m\circ}]^{-{1\over 2}}  \\
 \ \ \ \ \ \ \ \ \ \ \ \ \ \ \ \ \ \ \ \ \ \ \ \ \ \ \ \ \ \ \ \ \ \
\ (1+x)^{(m-3)} \ F^2(x)  \ dx\ .
 \end{array}
\end{equation}
Our 45 candidates were found in the $\sim 1500^{\tiny \sq}$ (square degrees) 
field of the 2dFGRS. With 
a redshift interval $[0.001, 0.108]$ and $m=3$, the local fequency is $f_\circ \sim 
9400 \ h^3 \ {\mathrm {Gpc}^{-3}}$. For $h=0.7$  we get 
$f_\circ \sim 3000 \ {\mathrm {Gpc}^{-3}}$. \\

We may evaluate the 
number of interactivated 
galaxies in a survey limited at $z_l \sim 3$. For the ``concordance" model and $m=3$ 
we deduce a 
number of 80 interactivated pairs by square degree ($\sim 80^{{\tiny  \ \sq} ^{-1}}$).  \\

In fact it is presumable that the comoving density
of interactivated pairs does increase with redshift, i.e. that the local real 
density $n(z_c)$ increases more steeply than $(1+z_c)^3$, leading to an underestimation
of $n$. Le F\`evre \& al (\cite{Lefev00}) derive a merger fraction of galaxies 
increasing
as $\propto (1+z_c)^m$ with $m={3.2\pm0.6}$.  Lavery \& al. (\cite {Lav04}) deduce from 
collisional ring galaxies in HST deep field a galaxy interaction/merger 
rate with $m = {5.2\pm0.7}$ or even steeper.
With $ m=5$, the number of expected pairs by square degree up to $z_c=3$ climbs to
$  \sim 700^{{\tiny \ \sq} ^{-1}}$, a comfortable density for MOS.
A $12^{{\tiny \ \sq}}$ survey would then supply
1 000 interactivated pairs, if $m=3$. Some 10 000 pairs would be a foreseeable target
in $100^{{\tiny \ \sq}}$ and the total number of interactivated galaxies
 in a whole sky survey could be more than $10^7$ if $m>4$.

\subsection{Inhomogeneities}
Our universe is no longer the realisation of an FL model. The presence of 
inhomogeneities modifies the $\theta_\circ \longleftrightarrow z_c$ relation. 
This problem is  difficult  to solve mathematically. 
It was  investigated long ago  
(Dashveski \& Zeldovich \cite {Zeldo65}, Dyer \& Roeder \cite {Dyer72}).
Hadrovi\'c \& Binney (\cite{Hadro97})   used the methods of gravitational lensing 
to measure the involved biases. They derived a bias of $-0.17\pm0.4$ on $q_\circ$,
and showed that larger objects yield to smaller errors. Demianski et al. (\cite{Demia03}) 
derive exact solutions of $\theta_\circ(z)$ 
for some cases of locally inhomogeneous
universes with a nonzero cosmological constant and approximate solutions for $z<10$.
We note from those previous works that the size of our standard of length
($\sim 100$ kpc) would make our method less sensitive to inhomogeneities
 than would parsec size ultra-compact radio sources.\\
 
We also note that carrying out our method on real pairs of interactivated galaxies presupposes 
acquiring a wide-field imaging 
of those objects and then detecting all the possibly intervening galaxies or clusters 
close to the lines of sight. 
It would then be easy to exclude the most perturbed lines of sight and to 
limit the sample to regular directions of intervening space.

\subsection{Observing}
Interactivated galaxies with a mean projected separation above 100 kpc
have (Sect. 3) a mean angular separation $\theta_\circ > 10 \arcsec$  over the 
whole range of $z$, and then the measure of $\theta_\circ$ will not add a significant dispersion
in the data (the main dispersion remaining that of linear separation).
The accuracy of measuring redshift $z$
 is not a problem for those strong emission-line objects
even with low dispersion spectroscopy 
(always compared to the intrinsic dispersion in the $\theta_\circ$ dimension).\\

The selection of primary interactivating pairs of galaxies seems achievable
by wide-field imaging. Candidates may be selected by  colours, magnitudes,
angular separations, and morphology: the first approach of the two partners generally 
preserves a much simpler geometry  for both of them than does the pre-merging 
second perigalacticon. 
Then a long or multi-slit spectrography with low dispersion (and low signal-to-noise ratio) 
would be enough to 
characterise and classify the 
starbursts and measure the redshifts. Integral field spectroscopy 
could be used -- via its potentiality to supply velocity fields -- to implement the classification 
criteria.\\

The main difficulty in running this program is obviously the faintness of those sources  
meant for spectroscopy with today's telescopes. Without K-correction or extinction the distance modulus is
supplied by the ``luminosity distance'' $d_L = d (1+z) $ 
(Mineur \cite{Min33}, Robertson, \cite{Robert38}):
$m-M = -5 + 5 \log d_L$. The brightest 
members of the 45 2dFGRS pairs used in our "real sample" would reach 
$V = 26$ if located at $z=3$ in a concordance model (and close to $V=28$ for the mean
of luminosities). But if we look at the distribution of 2dFQRS redshifts,  only 21\%
have $z>2.$ and  4\% $z>2.5$. The bulk of objects is centred on $z=1.6$, for which 
the V magnitudes would be 24.6 for the brightest ones and 26 for the mean of luminosity.
The rejection (at any $z$) of less luminous objects could be an operational
criterion.  \\

If K-correction and  -- mainly intrinsic -- extinction increase the 
above estimations, those
two effects presumably are over-compensated by the increase of the intensities of starbursts
with redshift: more gas in galaxies at remote times and the Schmidt law (Schmidt, 
\cite {Schmi59}) linking SFR to the  the density of gas. 
As a matter of fact and even if they are mainly concerned with the ``secondary''
-- pre-merger -- starburst, many  approaches in several wavelength ranges 
(see e. g. Mihos \& Hernquist, \cite {Mihos94}, Steidel \& al, \cite {Steid99}, 
Elbaz, \cite {Elbaz04}) measure  a rapid increase of a factor 
$\sim10$ (even without extinction correction) in the general SFR when looking backward in time from $z=0$ to $z \approx 1$ 
followed by a  quasi-constant rate up to $z>3$.\\

Restricting the selection of candidates to balanced pairs 
-- e.g. $\left | B_1 - B_2 \right | < 1$ -- could also be a means to favour strong starbursts.
Another fact could help build feasibility in the future: 
due to its observational selection the 2QZ survey is, as already mentioned, very poor in $2<z<3$ 
objects and concentrated around $z\approx 1.5 \pm 0.7$. In the real samples of 
interactivated galaxies, we may expect a  distribution of redshifts that is less 
vanishing. Present uncertainties on that evolution mean that
 we do not try to further compute the foreseeable 
distribution in $z$  of a real  sample of interactivated galaxies, but we do note that, 
 in conjunction with the increase of starbursts luminosities with $z$,  a high value of
  $m$  index
or a distribution of $z$ simply that is flatter than for 2QZ
would make the $\theta_\circ \longleftrightarrow z_c$ relation 
 more sensitive  to cosmological parameters, peculiar to $\Omega_{\Lambda\circ}$, but 
 with the drawback of an increase in the fraction of faint high z objects. \\

With the  foreseeable progress in the interactivation models, classifying 
diagnostics could be deduced. In each class the dispersion in linear (and projected) 
separations is expected to be lower, and 
each sub-sample could supply independent estimations of  $\Omega_{i\circ}$ thereby
 giving both a test and 
more accuracy.\\

Finally the method could perhaps be applied to much brighter objects like interactivated pairs of 
Seyferts, if it could be established that they also have a characteristic distance distribution.

\section{Conclusion}
We studied a new method of observational cosmology using the angular size 
versus redshift relation and ``primary
interactivation" of pairs of galaxies as a natural generator of yardsticks. 
The properties of the population was estimated from the 2dFGRS. The
number of those interactivated sources in the observable universe is much more than needed.
 Monte Carlo 
simulations show that an accuracy of $\pm 0.1$ on 
$\Omega_{m\circ}$ and $\Omega_{\Lambda\circ}$ seems feasible with a $\sim 10^{\tiny \sq}$ survey. 
Reaching $\pm 0.01$ would imply a much wider survey and additional tests against bias. 
The method could also be used for constraining other free parameters of 
non-vacuum dark energy, quintessence, or  other modified FL cosmologies.
The main problem seems to be the faintness of remote sources for spectroscopy 
with today's optical telescopes.

\begin{acknowledgements}
We are very grateful to L. Delaye and A. P\'epin for a simplified modelisation of 
the orbital and starburst parameters of 45 candidates extracted from the 2dFGRS. 
Many thanks to V. Springel and L. Hernquist  (\cite {Spri05}) for sending us the output tables of 
their synthetic interactivation and to the referee for all her/his pertinent comments.
\end{acknowledgements}

\end{document}